\documentclass[prd,twocolumn,aps,epsbox,groupedaddress,eqsecnum,notitlepage,nofootinbib]{revtex4}
\usepackage[dvipdf]{graphicx}
\usepackage{epstopdf}
\usepackage{amsmath}
\usepackage{slashed}
\usepackage{amssymb}
\usepackage{enumerate, color}
\usepackage{subfigure}
\usepackage{tabularx}
\usepackage{epstopdf}
\usepackage{lipsum}
\usepackage{color}
\usepackage[utf8]{inputenc}
\newcommand{\be}{\begin{equation}}
\newcommand{\ee}{\end{equation}}
\newcommand{\ben}{\begin{eqnarray}}
\newcommand{\een}{\end{eqnarray}}
\newcommand{\bes}{\begin{subequations}}
\newcommand{\ees}{\end{subequations}}

\begin{document} 
%%%%%%%%%%%%%%%%%%%%%%%%%%%%%%%%%%%%%%%%%%%%%%%%%%%%%%%%%%
%%%%%%%%%%%%%%%%%%%%%%%%%%%%%%%%%%%%%%%%%%%%%%%%%%%%%%%%%%
\title{Fermions in the presence of topological structures under geometric constrictions}
\author{D. Bazeia$^{1}$}
\author{A. Mohammadi$^{2}$}
\author{D. C. Moreira$^{3}$}
\affiliation{$^1$Departamento de F\'isica, Universidade Federal da Para\'iba, 58051-970, Jo\~ao Pessoa, PB, Brazil}
\affiliation{$^2$Departamento de F\'isica, Universidade Federal de Pernambuco, 50670-901, Recife, PE, Brazil}
\affiliation{$^3$Faculdade Uninassau Petrolina, 56308-210, Petrolina, PE, Brazil}
%%%%%%%%%%%%%%%%%%%%%%%%%%%%%%%%%%%%%%%%%%%%%%%%%%%%%%%%%%
%%%%%%%%%%%%%%%%%%%%%%%%%%%%%%%%%%%%%%%%%%%%%%%%%%%%%%%%%%

\begin{abstract}
In this work we study modifications of the spectrum of fermions interacting with kinklike structures in two-dimensional spacetime. We consider the Yukawa coupling between fermions and scalar fields that engender nontrivial internal structure and investigate how the fermion spectra change in terms of the parameters that control the kinklike configuration and the Yukawa coupling. We consider models that allow the internal structure of the kinklike solution to respond to the presence of a geometrical constriction, and show the fermion spectra may also appear directly affected by the constriction. The main results are of current interest and may be used to propose the construction of electronic devices capable of engendering new effects at the nanometric scale.

\bigskip

\end{abstract}

%\pacs{04.50.-h, 11.27.+d}

\maketitle
%%%%%%%%%%%%%%%%%%%%%%%%%%%%%%%%%%%%%%%%%%%%%%%%%%%%%%%%%%%%%%%%%%%%%%%%%%%%%%%%%%%%%%%%%%%%%%%%%%%%%%%%%%%%%%%%%%%%
\section{Introduction}
The study of localized structures with a topological character has several implications in Physics. In high energy physics, in particular, we can list a group of topological solutions whose stability also depends on the dimension of the background spacetime \cite{manton2004topological,vachaspati2006kinks}. As one knows, there are kinks in the line, vortices in the plane and magnetic monopoles in three spatial dimensions. In the simplest case, in $(1,1)$ spacetime dimensions kinks are generated by real scalar fields and can be used in a diversity of applications; see, e.g., \cite{k1,k2,k3,k4,A1,A2} and references therein. In the presence of two real scalar fields, in particular, there are several interesting possibilities, as the ones considered in \cite{T1,T2}, for instance, where the second field may contribute to describe internal structure \cite{DB}. More recently, several works have appeared, focused on the study of the asymptotic behavior of kinks to analyse what sort of effects may appear in inter-kink interactions \cite{christov2019kink,manton2019forces,bazeia2018analytical,christov2019long,khare2019family} and in the scattering of kinks \cite{gani2019multi,bazeia2019scattering,belendryasova2019scattering,bazeia2018scattering,alonso2019asymmetric,bazeiagomes,AM20}. 

The presence of kinks has inspired several other possibilities, a recent one considering the case of scalar fields in the presence of impurity \cite{in0,in00}. The subject is of current interest and in \cite{in1}, for instance, the authors found a family of impurity models such that the self-dual sector is exactly solvable, for any spatial distribution of the impurity, both in the topologically trivial case and for kink configurations. Moreover, in \cite{in2} an interesting procedure is described, in which a first order equation for static kink is solved iteratively, also leading to the construction of exact solutions. Another line of investigation concerns the use of kinks and domain walls  to describe specific properties of magnetic materials. One issue of current interest here is the possibility to construct high performance memory devices \cite{science} and their interplay with skyrmions \cite{nature1,nature2}. 

  Another subject that engenders important applications of kinks, in which we will pay closer attention in this work, is related to the Jackiw-Rebbi (JR) model \cite{jackiw1976solitons}. The issue here concerns the use of kinks as background to study the fermion spectrum. In the JR model one can show that, when introducing topological solutions in the Dirac equation, the symmetry breaking required by the scalar potential to construct the topological structure implies the fractionalisation of the fermion number due to the nontrivial background described by the scalar field. Some standard works on this subject are in refs. \cite{su1979solitons,jackiw1981solitons,niemi1986fermion,heeger1988solitons,chamon} and there are recent developments in \cite{yang2019soliton,guilarte2019fractional,alonso2019soliton} and in references therein. Since in the context of fermion-kink interactions the properties of the fermion field depend on how the scalar field behaves, it is natural that topological solutions with different properties, either in the asymptotic behavior or in the internal structure, induce changes in the distribution of the fermion modes, which has been recently explored in some specific setups in Refs. \cite{amado2017coupled,bazeia2017fermionic,bazeia2019fermion,azadeh}. 

Motivated by the possibility to study models that obey first-order equations with solutions saturating the Bogomol'nyi-Prasad-Sommerfield bound \cite{Bogomolny:406760,prasad1975exact}, a new class of kinklike structures where the scalar field acquires non-trivial internal structure has been proposed \cite{bazeia2020geometrically}. The underlying mechanism is assembled with two real scalar fields that interact through a potential and the modification of the kinematics of the first field, while the second field develops standard evolution. The general procedure engenders first-order differential equations that can be solved analytically. Compared with the JR model, in the novel model the symmetry is enhanced to accommodate the second field, which is adjusted to have a kink profile that induces the appearance of distinct behavior on the first field, simulating a geometric constriction as the one investigated before in \cite{rapid}. The richness of the procedure has suggested that we further explore how the internal structures emerging from the kinklike configuration may affect the behavior of the fermion spectrum when the scalar field interacts through the Yukawa coupling. The idea here is to build a connection with the experimental result described in Ref. \cite{prlpolar}, where an electric current is used to modify the polarity of a localized structure nested inside a magnetic nanowire.

To focus on this issue, we organise the work as follows. In Sec. II we present the scalar and fermion Lagrangian densities that compose the model under study and discuss the formation of kinks with internal structure and the equations of motion of the fermion in the background of the scalar configurations. There we also investigate some specific models within the above context, with the focus on the spectrum of fermions, and in Sec. III we end up the work with some conclusions and comments. In particular, we discuss possibility to construct electronic devices at the nanometric scale, suggested to identify the new effect captured by the presence of geometric constrictions in nanowires. 

%%%%%%%%%%%%%%%%%%%%%%%%%%%%%%%%%%%%%%%%%%%%%%%%%%%%%%%%%%
%%%%%%%%%%%%%%%%%%%%%%%%%%%%%%%%%%%%%%%%%%%%%%%%%%%%%%%%%%
\section{Fermion-kink system}

Let us now analyse a model with a fermion field interacting with a bosonic one with a kink profile {\it via} Yukawa coupling, described by the Lagrangian $\mathcal{L}=\mathcal{L}_f+\mathcal{L}_b$, where
\bes\ben
\mathcal{L}_f &=&  \frac{1}{2}\bar{\psi}\,i\gamma^\mu{\partial_\mu} \psi - g\phi\,\bar{\psi}\psi,\label{lf} \\
\mathcal{L}_b &=&  \frac{1}{2}f(\chi)\partial_\mu\phi\partial^\mu\phi+\frac12\partial_\mu\chi\partial^\mu\chi-V(\phi,\chi).\;\;\;\label{lb}
\een\ees
In the above system, $\psi$ stands for the Dirac field and the scalar field $\phi$ is used to describe a background kinklike solution coming from the Lagrangian density in \eqref{lb}, where we have two interacting real scalar fields, the second scalar field denoted by $\chi$. Here, $f(\chi)$ is in principle an arbitrary nonnegative function, and the potential has the form
\be\label{pot}
V(\phi,\chi)=\frac{1}{2f(\chi)}(1-\phi^2)^2+\frac{1}{2}\alpha^2(1-\chi^2)^2.
\ee
The bosonic model was used recently in \cite{bazeia2020geometrically} to generate kinklike solutions with interesting internal structures. We consider that $\alpha$ is a real and nonnegative parameter, and for $\chi=\pm1$ and $f(\pm1)=1$ we recover the original JR model \cite{jackiw1976solitons}. In this sense, in this work the $\chi$ field will be used as an independent field: it will act as a geometric constriction, contributing to distort the localized structure described by the $\phi$ field, which will then modify the fermion behavior due to the Yukawa coupling that appear in \eqref{lf}. Note that dimensionless fields, coupling constants and coordinates are been used in the present work.

The interesting point about this model is that the field $\chi$ acts independently and entraps a non-trivial kink configuration for the scalar field $\phi$, while the $\alpha$-parameter controls the geometry of the kink by making the entrapment more or less expressive. The net effect of this arrangement is that one of the scalar solutions acquires internal structures that can, in turn, reflect on aspects of the fermion spectrum in the above fermion-kink model. Here we consider three distinct models with $f(\chi)$ given by
\begin{align}\label{chimodels}
    f(\chi) = 
    \begin{cases}
      1/\chi^2,\\
      1/\cos^2(n \pi \chi),\\
      1/\sin^2((2k+1) \pi \chi)
    \end{cases}
\end{align}
with $n=1,2,3...$ and $k=0,1,2,...$ \cite{bazeia2020geometrically}; notice that $f(\chi=\pm1)=1$ in all the three cases.

\subsection{Bosonic System}

The bosonic model which appears in Eq. \eqref{lb} was first investigated in \cite{bazeia2020geometrically}, so here we briefly review some results there obtained. In the case of static solutions, the equations of motion for $\chi$ and $\phi$ are
\ben
\frac{d^2\chi}{dx^2}-\frac12\frac{df}{d\chi}\left(\frac{d\phi}{dx}\right)^2=
\frac{\partial V}{\partial\chi},\label{eom1}
\\
\frac{d}{dx}\left(f(\chi)\frac{d\phi}{dx}\right)=\frac{\partial V}{\partial \phi},\label{eom2}
\een 
and the energy density has the form
\be  
\rho(x)=\frac12\left(\frac{d\chi}{dx}\right)^2+ \frac12 f(\chi)\left(\frac{d\phi}{dx}\right)^2+ V(\phi,\chi).
\ee
As shown in \cite{bazeia2020geometrically}, the choice
\be 
V(\phi,\chi)=\frac{W^2_\phi}{2f(\chi)}+\frac{W^2_\chi}{2},
\ee
where $W=W(\phi,\chi)$, and $W_\phi=\partial W/\partial\phi$ and $W_\chi=\partial W/\partial\chi$, allows that we write the first order equations
\be 
\frac{d\chi}{dx}=\pm W_\chi(\phi,\chi),\\
\;\;\;\;\;
\frac{d\phi}{dx}=\pm\frac{W_\phi(\phi,\chi)}{f(\chi)},
\ee
which turn the energy density into a surface term, with the energy minimized to the value $E=|\Delta W|$, where $\Delta W=W(\phi(\infty),\chi(\infty))-W(\phi(-\infty), \chi(-\infty))$. The above results are important because solutions of the first order equations solve the equations of motion and are minimum energy configurations. Moreover, from the first order equations above, we see that when $W(\phi,\chi)$ can be written as
$W(\phi,\chi)=g(\phi)+h(\chi)$, they become
\be  
\frac{d\chi}{dx}=\pm\frac{dh(\chi)}{d\chi},\;\;\;\;\;\frac{d\phi}{dx}=\pm\frac1{f(\chi)}\frac{dg(\phi)}{d\phi}.
\ee
This means that the $\chi$ field is independent and can be solved independently. However, it acts to modify the behavior of $\phi$ under the presence of the function $f(\chi)$. The bosonic model then has the power to describe kinklike structures in the presence of geometrical constrictions, depending on the form of the functions $f(\chi)$, $g(\phi)$ and $h(\chi)$ that we choose to describe the model. This was stressed before in Ref.  \cite{bazeia2020geometrically}, where the choices displayed in Eqs. \eqref{pot} and \eqref{chimodels} were considered.

In order to further illustrate this, let us use the potential in Eq. \eqref{pot} to write the above first order equations as, taking the positive sign, for simplicity,
\be  
\frac{d\chi}{dx}=\alpha(1-\chi^2),\;\;\;\;\;\frac{d\phi}{dx}=\frac{(1-\phi^2)}{f(\chi)}.
\ee
The equation for $\chi$ is well-known and can solved immediately; it gives the solution $\chi(x)=\tanh(\alpha(x-x_0))$, where ${ x_0}$ is an integration constant which informs us the position of the center of the kinklike configuration, such that $\chi(x_0)=0$. Due to translational invariance, it is usually taken to be zero, and we will also take $x_0=0$, for simplicity. To solve the equation for $\phi$, we change variable from $x\to y(x)$ such that $dy/dx=1/f(\chi)$. This allows that we write the first order equation for $\phi$ in the much simpler form
\be  
\frac{d\phi}{dy}=1-\phi^2,
\ee 
which has the solution $\phi(y)=\tanh(y-y_0)$, such that $\phi(y_0)=0$. We now consider the model with $f(\chi)=1/\chi^2$, the first possibility that appears in Eq. \eqref{chimodels}. In this case we have ${dy}/{dx}=\tanh^2(\alpha x),$
which implies that $y(x)=x-(1/\alpha) \tanh(\alpha x),$ with $y(0)=0$. This gives
\be\label{phiX}
\phi(x)=\tanh\left(x-\frac{1}{\alpha}\tanh\alpha x-y_0\right),
\ee
such that when $x\to 0$ we have $\phi(0)=-\tanh(y_0)$, showing that the center of the kinklike structure described by $\phi$ may be shifted along the vertical axis. The other two cases displayed in Eq. \eqref{chimodels} works similarly.

In the present work, we want to investigate the fermion behavior guided by the Lagrangian in Eq. \eqref{lf}, to compare the results with the standard results obtained before in the JR model \cite{jackiw1976solitons}. For this reason, we further impose that $y_0$ in Eq. \eqref{phiX} vanishes, such that the kinklike configuration described by $\phi$ now engenders the behavior $\phi(-x)=-\phi(x)$ which is present in the JR model. The same behavior must be considered for the other two choices of $f(\chi)$, as shown in Eq. \eqref{chimodels}.

\subsection{Fermionic System}

We now turn attention to the fermion system. One defines as {\it ansatz} for the fermion field $\psi(x,t)=e^{-iEt} \psi(x)$, with $\psi(x)=\left(\psi_+(x),~\psi_-(x)\right)^T$, and uses the representation $\left(\gamma^0,\gamma^1,\gamma^5\right)=\left(\sigma_1, i\sigma_3,\sigma_2\right)$ for the Dirac gamma matrices, the equations of motion for the fermion components are given by
\bes\label{fermion2k}\ben
     E\ \psi_+ + \psi_-' -2 g \,\phi\,\psi_- &=& 0,\\
     E\ \psi_- - \psi_+' -2 g \,\phi\,\psi_+ &=& 0,
\een\ees
and the decoupled Schr\"odinger-like equations become
\begin{equation}\label{decoupledsystem}
\left(-\frac{d^2}{dx^2}+U_{\pm}(x)\right)\psi_{\mp}=E^2\psi_{\mp},
\end{equation}
where $U_{\pm}(x)=\mp 2gd\phi/dx+4g^2\phi^2$, with $\phi=\phi(x)$. It is easy to find the fermion zero mode, given by
\begin{align}\label{zeromode}
\psi(x)=\mathcal N\left(
\begin{array}{c}
e^{-2g\int^x\,{\phi(x') dx'}}\\
0\\
\end{array}\right)
\end{align}
with normalization factor $\cal N$. Beyond that, the half-bound fermion energies separating the bound and continua spectra are at $E_{hb}=\pm 2 g \phi(x\to \infty)$ and since for all the three background models in \eqref{chimodels} one has $\phi(x\to \infty) \to 1$ (see below) we end up with $E_{hb}=\pm 2 g$.

The three models presented so far have parity symmetry meaning that, regardless of the value of the parameters $\alpha$, $n$ and $k$, the upper and lower components of the fermion field have opposite parities. Moreover, the models have charge and energy-reflection 
symmetries. From the latter follows that the fermion energy spectrum is symmetric with respect to the line $E=0$. The charge conjugation operator is representation dependent and with the representation adopted here is $\sigma_3$, together with the complex conjugation of the wave function. Moreover, the energy-reflection and parity operators are representation independent and are both $\gamma^1$, along with $E\to -E$ and $x\to -x$, respectively. 
%%%%%%%%%%%%%%%%%%%%%%%%%%%%%%%%%%%%%%%%%%%%%%%%%%%%%%%%%%
%%%%%%%%%%%%%%%%%%%%%%%%%%%%%%%%%%%%%%%%%%%%%%%%%%%%%%%%%%
\subsection{Specific Models}

We now concentrate on some specific bosonic models, which allow obtaining distinct kinklike structures that serve as background for the charged fermions. The main aim is to see how the internal structure of the kink may induce the presence of fermion bound states.

\subsubsection{First Model}
\begin{figure}[t!]
 \subfigure[$ $]{\includegraphics[width=0.8\linewidth]{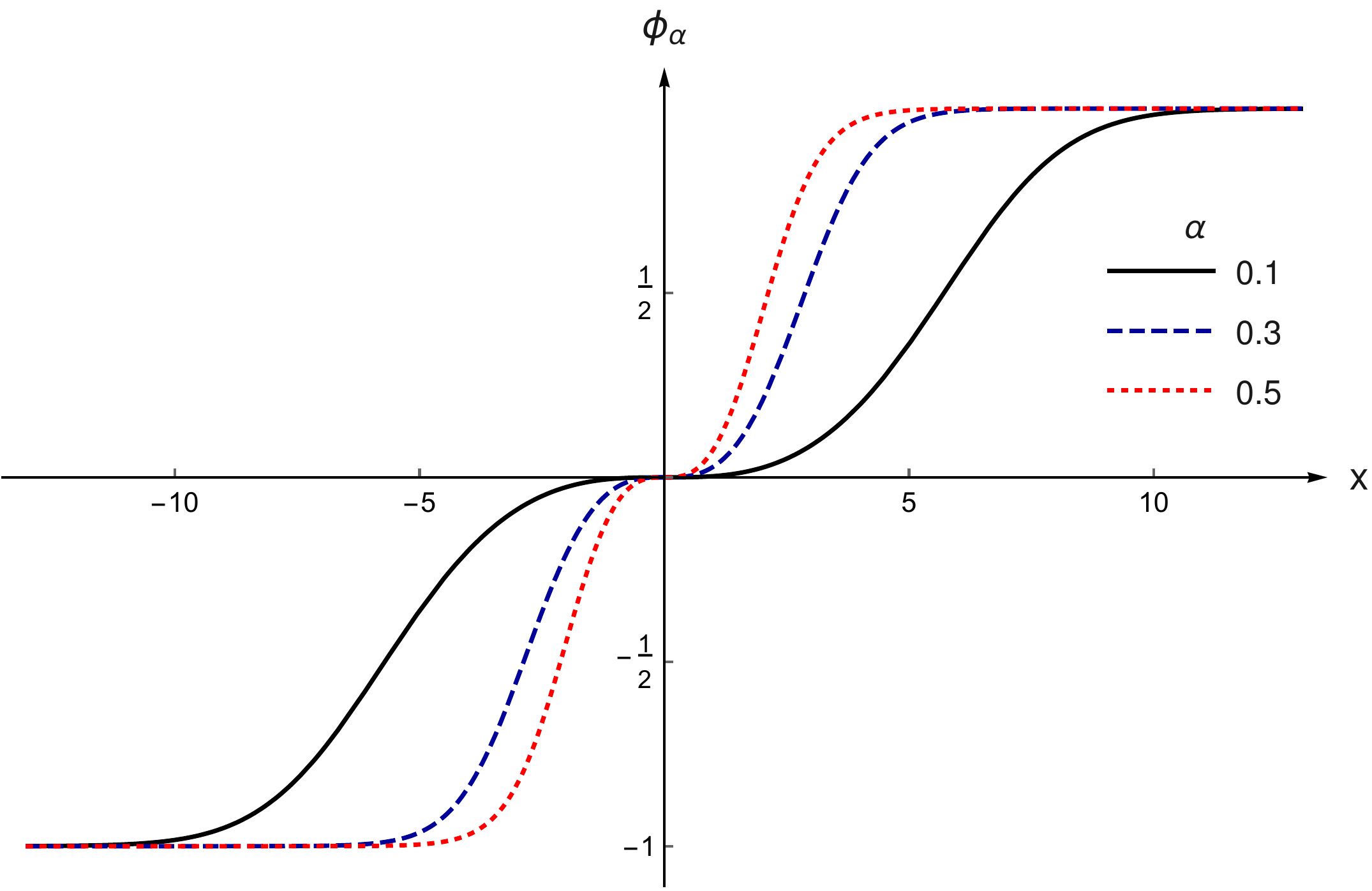}\label{fig1a}}\\
\subfigure[$ $]{\includegraphics[width=0.8\linewidth]{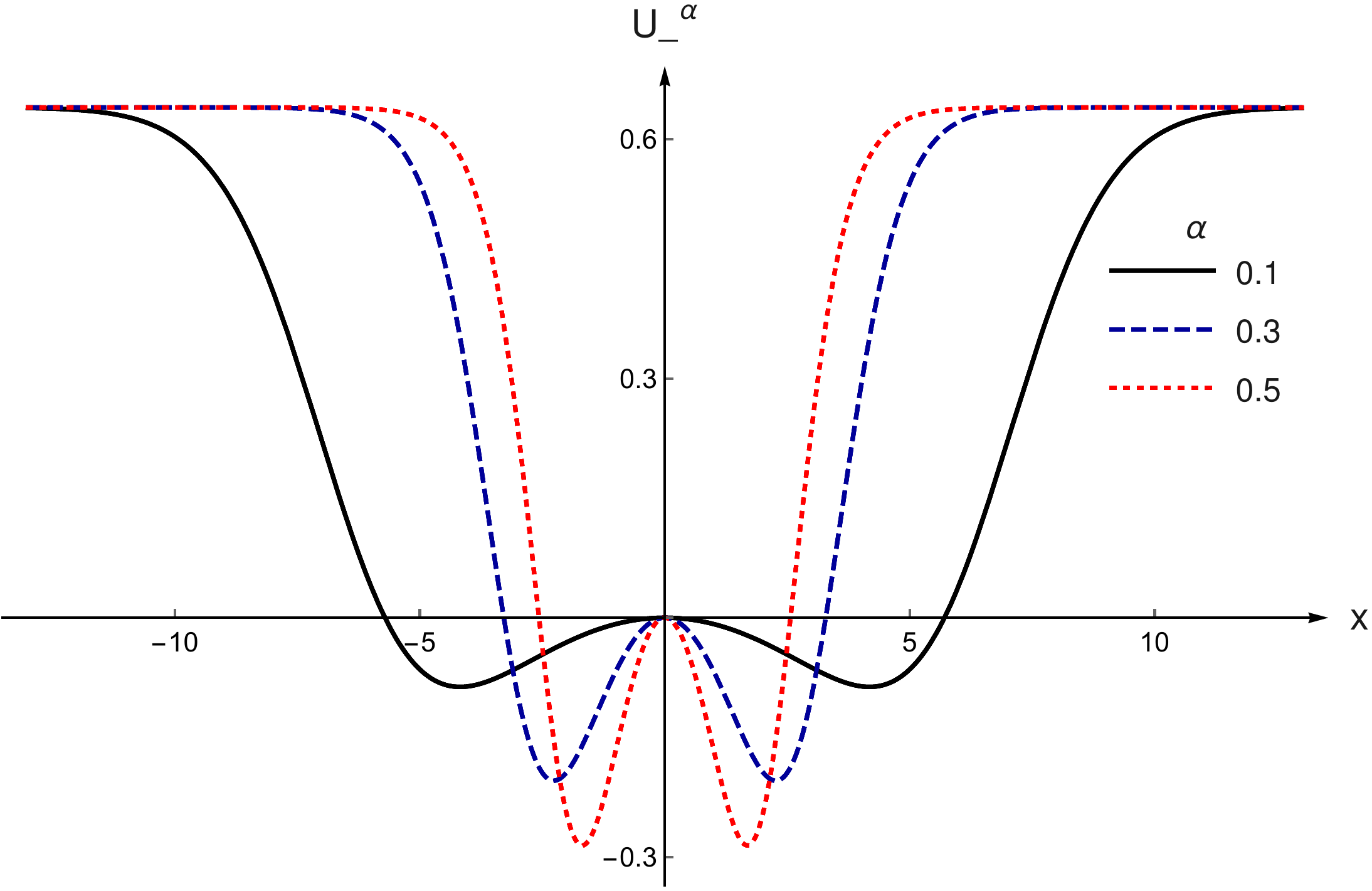}\label{fig1b}} 
  \caption{(a) Kink profile and (b) fermion potential for the first model, with $g=0.4$ and distinct values of $\alpha$.\label{fig1} }
  \end{figure}

First, let us consider the model with $f(\chi)=1/\chi^2$. The solution for $\phi(x)$ in this model is given by Eq. \eqref{phiX} with $y_0=0$. It has the form
\begin{equation} \label{kink-mod1}
\phi_{\alpha}(x)=\tanh \left(x-\frac1\alpha\tanh(\alpha x)\right),
\end{equation}
The solution is shown in Fig. \ref{fig1}(a) considering several values of $\alpha$. The plateau in the solution around zero diminishes as $\alpha$ increases. For $\alpha>0$ the slope of the kink at the center is zero and $\alpha$ can be used to control the size of the plateau in the kinklike configuration. This behavior was firstly noticed in \cite{bazeia2017fermionic}, and can be used in applications of practical interest.

\begin{figure}[t!]
 \subfigure[$~$]{\includegraphics[width=0.6\linewidth]{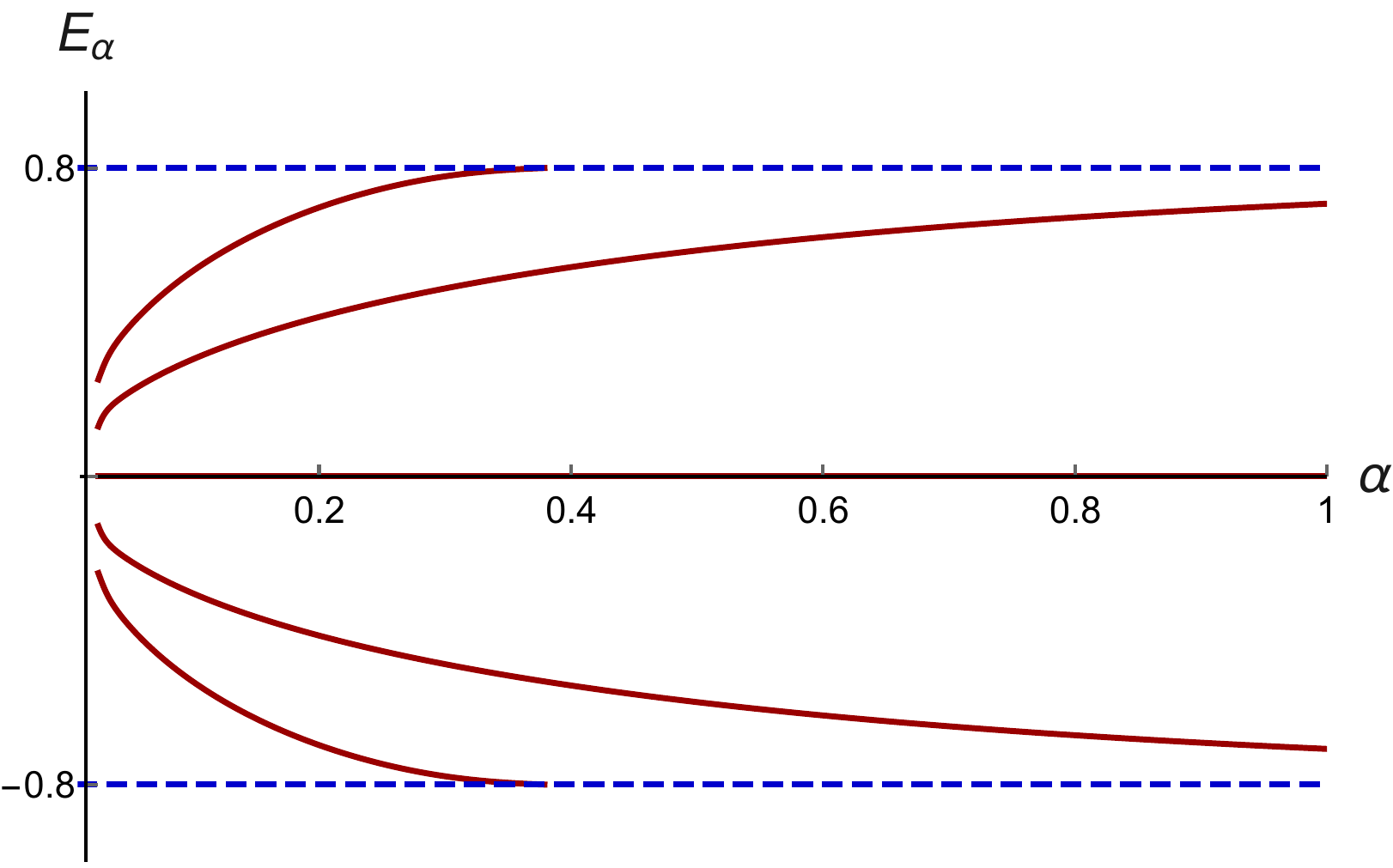}\label{fig2a}}\\ \hspace{2mm}
\subfigure[$~$]{\includegraphics[width=0.6\linewidth]{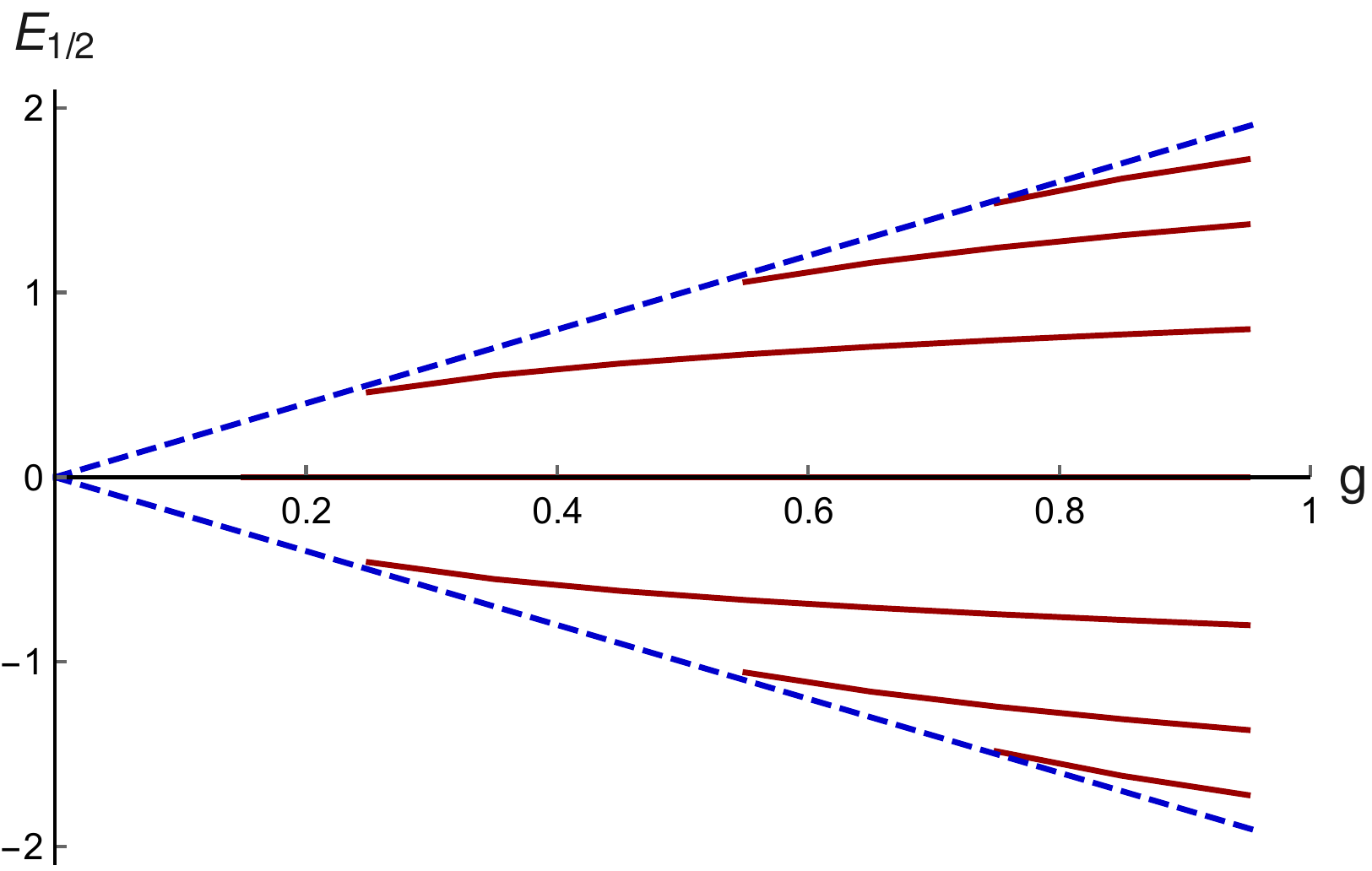}\label{fig2b}} 
  \caption{Fermion spectra in the first model, with (a) $g=0.4$ and continuous $\alpha$ and with (b) $\alpha=0.5$ and continuous $g$. In both cases, the dashed lines show the half-bound state energy.\label{fig2} }
  \end{figure}

By substituting the kink solution (\ref{kink-mod1}) as the background field $\phi(x)$ in the Eq. \eqref{decoupledsystem} one finds for the fermion potential
\be
U_{\pm}(x)=\pm 2\mathrm{sech}^2y(x)\tanh^2(\alpha x)+4\tanh^2y(x),
\ee
where $y(x)=x-(1/\alpha)\tanh(\alpha x)$. The potential $U_-$ which responds for the entrapment of the fermion bound states is shown in Fig. \ref{fig1}(b) for several values of $\alpha$. As the parameter $\alpha$ increases the depth of the fermion potential below zero increases leading to the higher absolute value of the fermion bound energy. This phenomenon can be seen in Fig. \ref{fig2}(a). There could also appear more bound states, however, this is not the case in this model in contrast with that similar model in \cite{bazeia2017fermionic}. The fermion bound states as a function of the coupling constant $g$ is shown in Fig. \ref{fig2}(b). As one can note, the only fermion bound state is the zero mode for small coupling $g$ but as the value of $g$ increases, more fermion bound states arise. The fermion spectrum is completely symmetric in positive and negative energies reflecting the energy-reflection symmetry discussed before. Finally, we have to have in mind that when dealing with models in fermion-kink setups one must take some care since for large values of the coupling constant the kink used as the background field may not be a good approximation.  

%%%%%%%%%%%%%%%%%%%%%%%%%%%%%%%%%%%%%%%%%%%%%%%%%%%%%%%%%%%%%%%%%%%%%%%%%%%%%%%%%%%%%%%%%%%%%%%%%%%%%%%%%%%%%%%%%%%%

%%%%%%%%%%%%%%%%%%%%%%%%%%%%%%%%%%%%%%%%%%%%%%%%%%%%%%%%%%%%%%%%%%%%%%%%%%%%%%%%%%%%%%%%%%%%%%%%%%%%%%%%%%%%%%%%%%%%
%%%%%%%%%%%%%%%%%%%%%%%%%%%%%%%%%%%%%%%%%%%%%%%%%%%%%%%%%%%%%%%%%%%%%%%%%%%%%%%%%%%%%%%%%%%%%%%%%%%%%%%%%%%%%%%%%%%%

  \subsubsection{Second Model}
  
\begin{figure}[t!]
 \subfigure[$~$]{\includegraphics[width=0.76\linewidth]{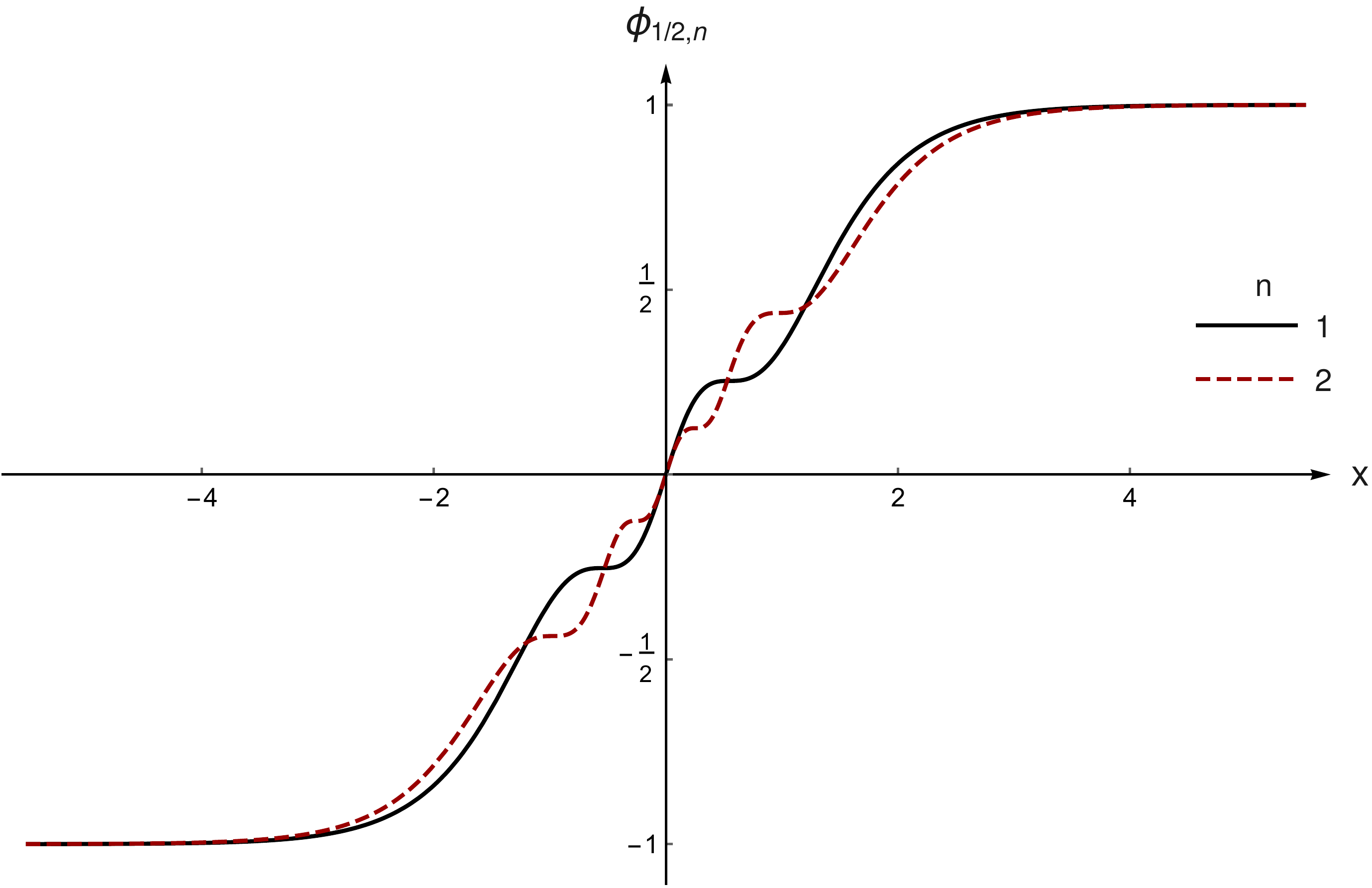}\label{fig3a}}\\ %\hspace{5mm}%\hfill or \hspace{5mm} or \hspace{0.3\textwidth}
\subfigure[$~$]{\includegraphics[width=0.76\linewidth]{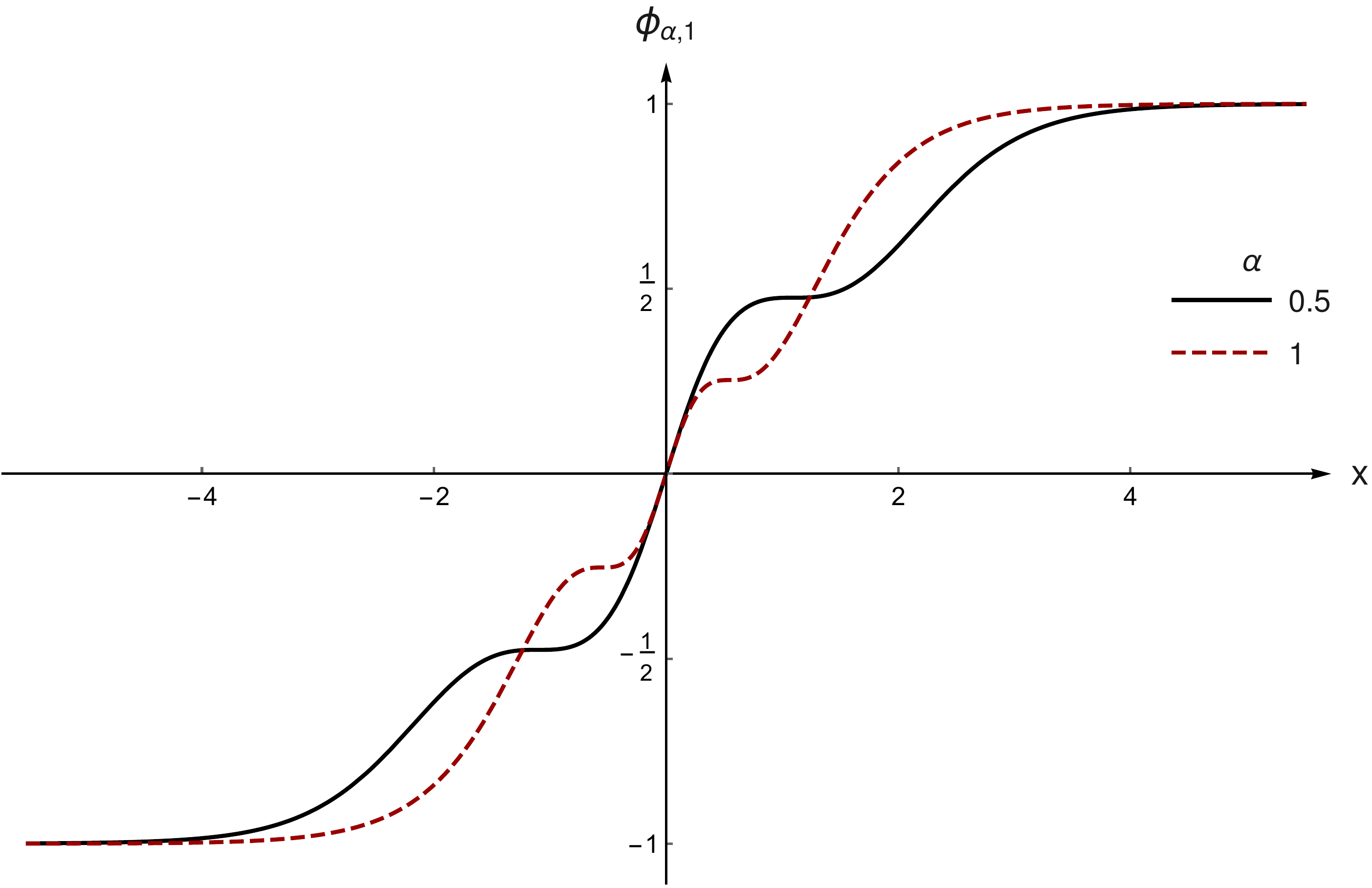}\label{fig3b}} 
  \caption{Kink profile for the second model, with $g=0.4$: (a) for $\alpha=0.5$ and $n=1,2$ and (b) for $n=1$ and $\alpha=0.5, 1$. \label{fig3} }
  \end{figure}

The second model we consider in this work is represented by $f(\chi)=1/\cos^2(n\pi\chi)$ and $n=1,2,3,...$, which yields the solution
\begin{equation}\label{kinkm2}
\phi_{\alpha, n}(x)=\tanh \frac{1}{2}\left(x+\frac{1}{2\alpha}\left(\text{Ci}\left(y_n^+\right)-\text{Ci}\left(y_n^-\right)\right)\right),
\end{equation}
where $\text{Ci}(y)$ denotes the cosine integral function and $y_n^{\pm}=2n\pi\left(1\pm \tanh(\alpha x)\right)$; see \cite{bazeia2020geometrically}. In Fig. \ref{fig3}(a,b) we display the profile of the solution \eqref{kinkm2} considering some values of $n$ and $\alpha$. We can observe that when increasing the values of $n$, more plateaux appear connecting the asymptotic limits of the solution, counting as $2n$. Besides that, increasing the value of the parameter $\alpha$ from zero leads to constrained structures concentrated around the center. We also notice that, the slope of the kink at the center is nonzero and equals unit independent of the values of the parameters $n$ and $\alpha$, differently from what occurs in the model presented in the previous subsection.

%%%%%%%%%%%%%%%%%%%%%%%%%%%%%%%%%%%%%%%%%%%%%%%%%%%%%%%%%%%%%%%%%%%%%%%%%%%%%%%%%%%%%%%%%%%%%%%%%%%%%%%%%%%%%%%%%%%%
%%%%%%%%%%%%%%%%%%%%%%%%%%%%%%%%%%%%%%%%%%%%%%%%%%%%%%%%%%%%%%%%%%%%%%%%%%%%%%%%%%%%%%%%%%%%%%%%%%%%%%%%%%%%%%%%%%%%
%%%%%%%%%%%%%%%%%%%%%%%%%%%%%%%%%%%%%%%%%%%%%%%%%%%%%%%%%%%%%%%%%%%%%%%%%%%%%%%%%%%%%%%%%%%%%%%%%%%%%%%%%%%%%%%%%%%%
\begin{figure}[t!]
 \subfigure[$~$]{\includegraphics[width=0.76\linewidth]{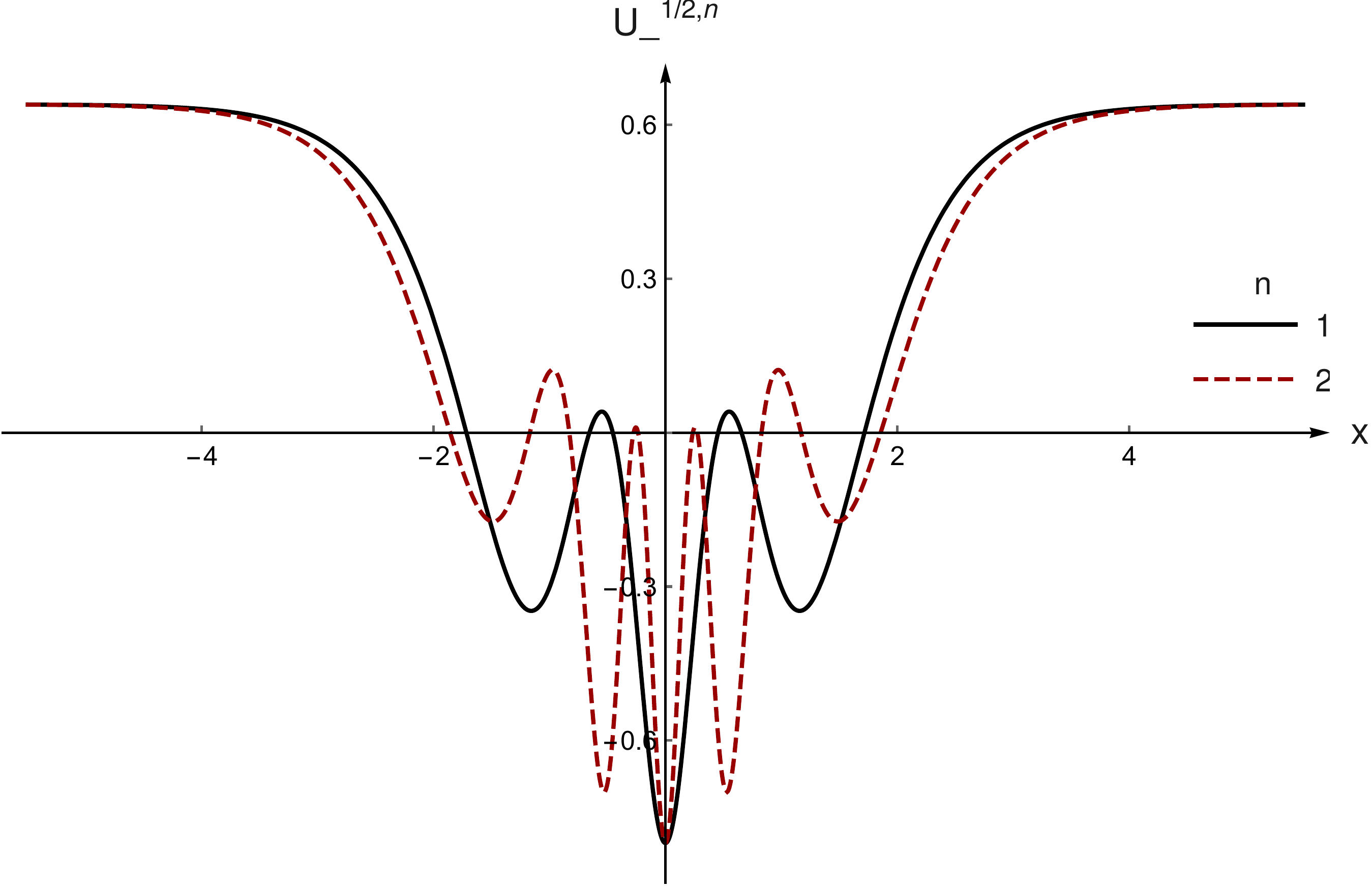}\label{fig4a}}\\ %\hspace{5mm}%\hfill or \hspace{5mm} or \hspace{0.3\textwidth}
\subfigure[$~$]{\includegraphics[width=0.76\linewidth]{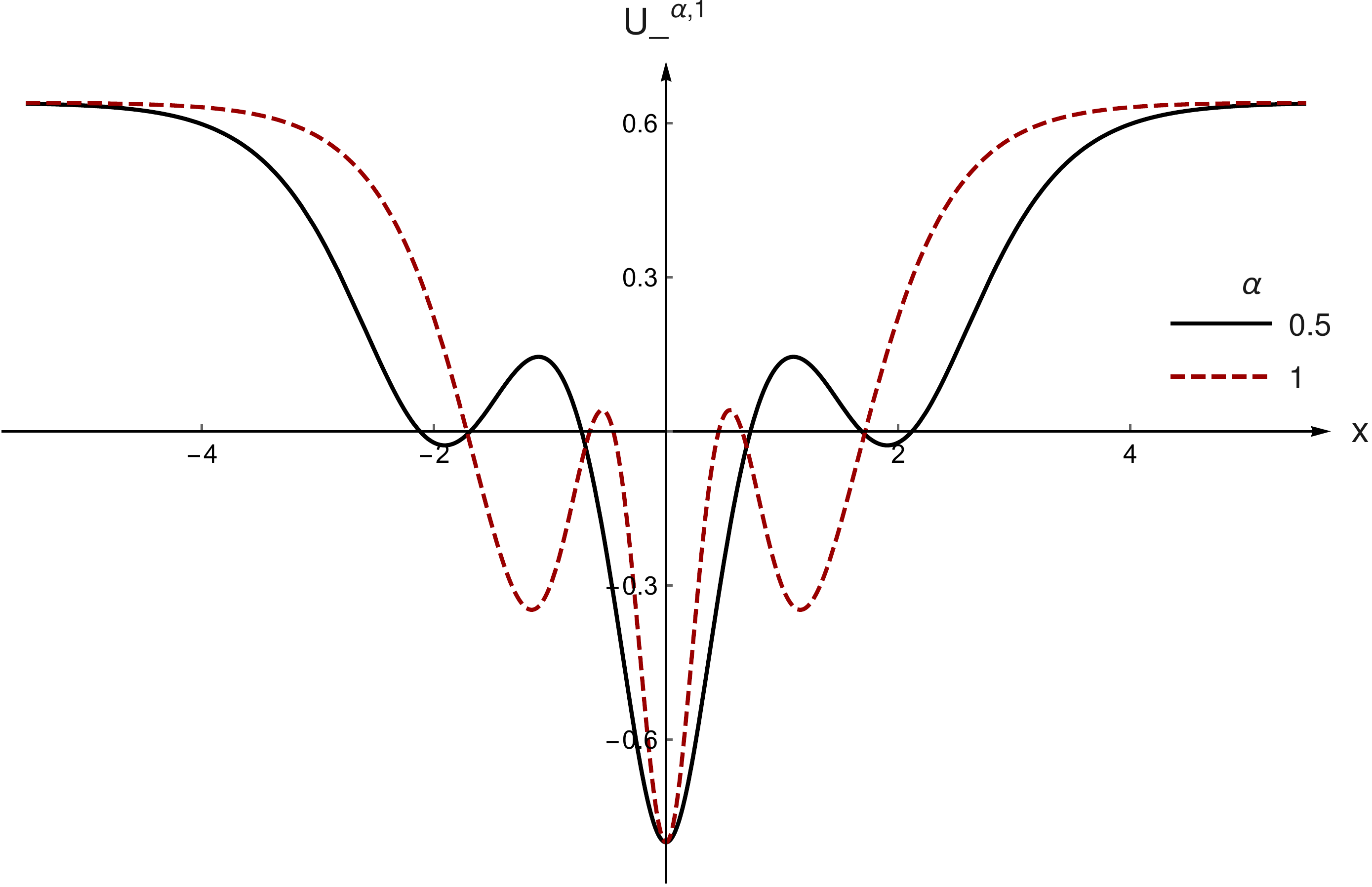}\label{fig4b}} 
  	\caption{Fermion potential for the second model, with $g=0.4$: (a) for $\alpha=0.5$ and $n=1,2$ and (b) for $n=1$ and $\alpha=0.5, 1$. \label{fig4} }
  \end{figure}
%%%%%%%%%%%%%%%%%%%%%%%%%%%%%%%%%%%%%%%%%%%%%%%%%%%%%%%%%%%%%%%%%%%%%%%%%%%%%%%%%%%%%%%%%%%%%%%%%%%%%%%%%%%%%%%%%%%%

\begin{figure}[t!]
 \subfigure[$~$]{\includegraphics[width=0.6\linewidth]{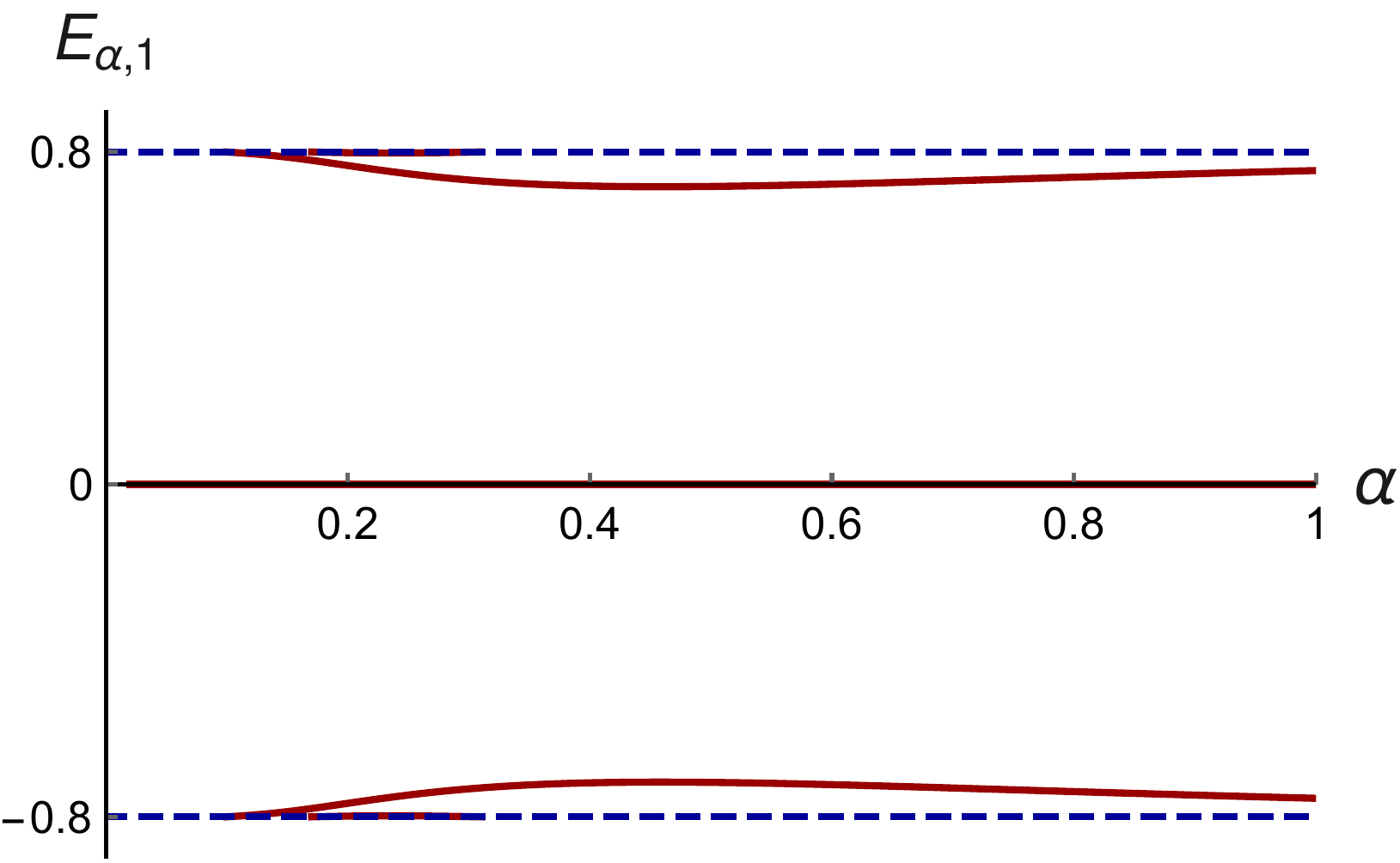}\label{fig5a}} \\%\hspace{2mm}%\hfill or \hspace{5mm} or \hspace{0.3\textwidth}
\subfigure[$~$]{\includegraphics[width=0.6\linewidth]{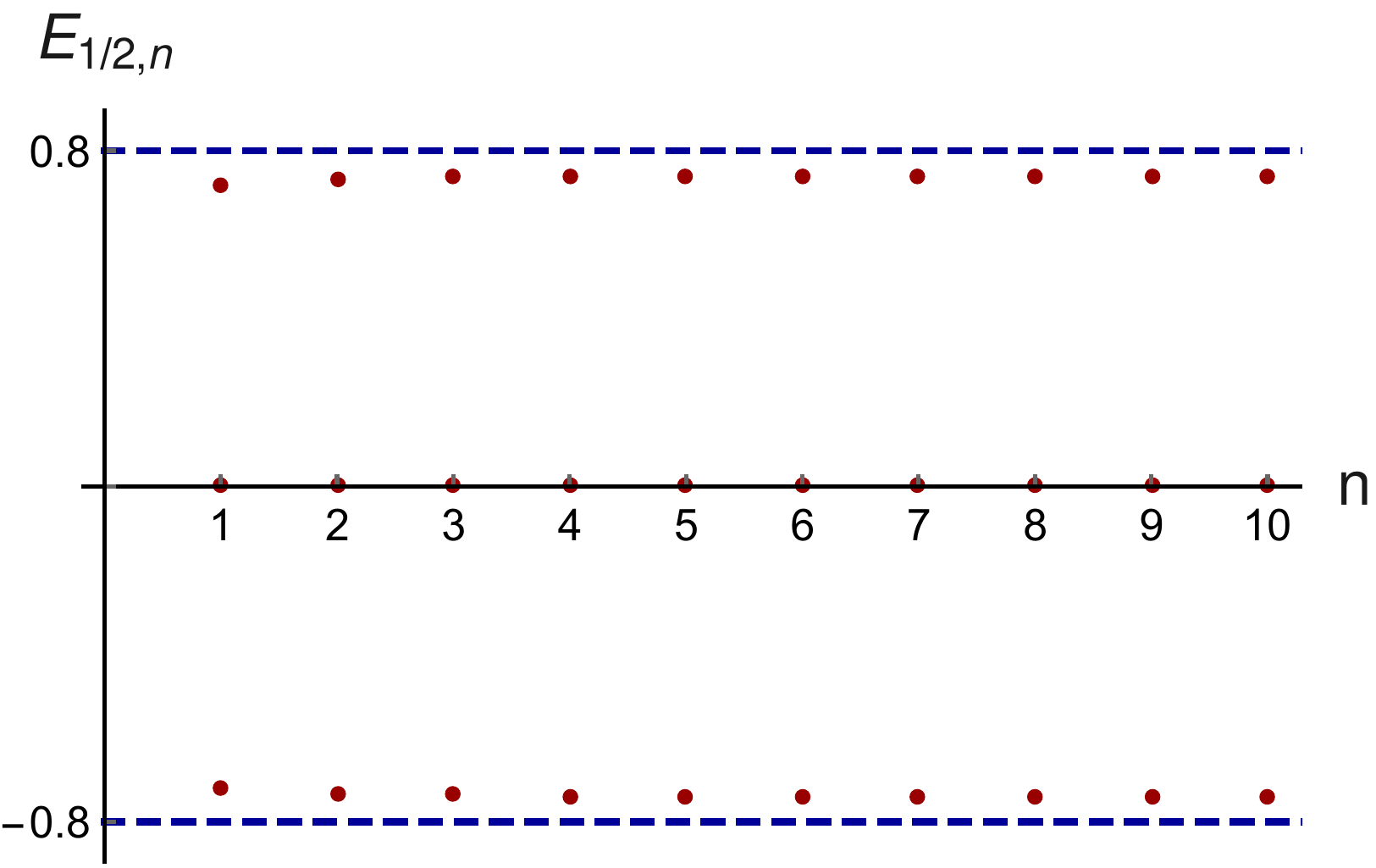}\label{fig5b}}\\ \hspace{2mm}%\hfill or \hspace{5mm} or \hspace{0.3\textwidth}
\subfigure[$~$]{\includegraphics[width=0.6\linewidth]{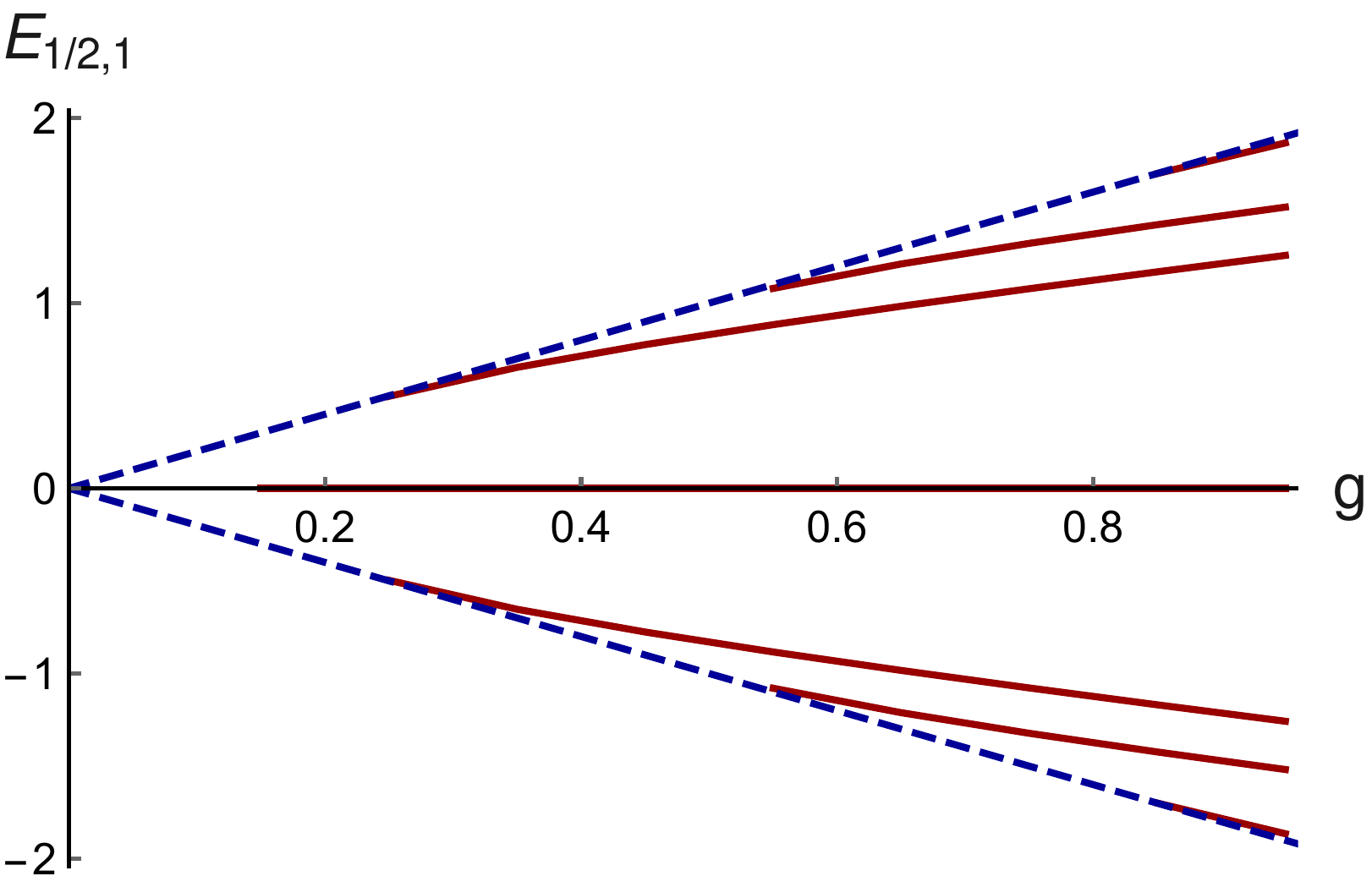}\label{fig5c}} %\hspace{5mm}
  	\caption{Fermion spectra for the second model, in terms of: (a) $\alpha$, for $g=0.4$ and $n=1$; (b) $n$, for $g=0.4$ and $\alpha=0.5$,
  	and (c) $g$, for $\alpha=0.5$ and $n=1$. \label{fig5}}
  \end{figure}

The fermion potential $U_{-}(x)$ is shown in Fig. \ref{fig4} for several values of the parameters $n$ and $\alpha$. In contrast with the potential presented for the previous model, the depth of the potential here is independent of the parameters of the model, making the fermion spectrum less sensitive to the variations of these parameters as one can see in Fig. \ref{fig5}, where the fermion bound spectrum is depicted in terms of the three parameters of the system, $n$, $\alpha$ and $g$. Again, the spectrum is symmetric around the line $E=0$, as expected. The dependence of the bound spectrum on $g$ is similar to the one arising in the previous model. 

\subsubsection{Third model}

\begin{figure}[t!]
 \subfigure[$~$]{\includegraphics[width=0.6\linewidth]{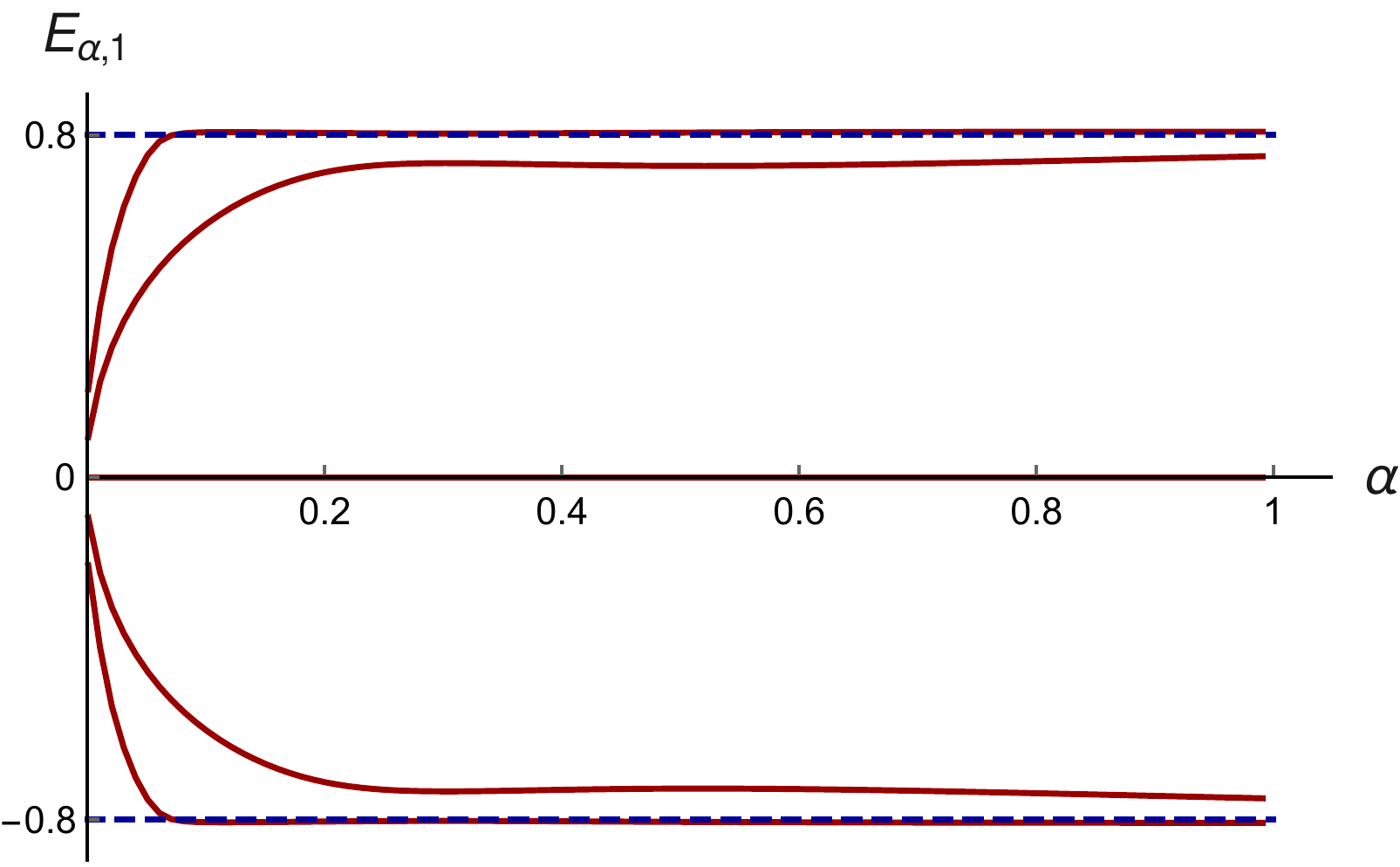}\label{fig6a}}\\ %\hspace{2mm}%\hfill or \hspace{5mm} or \hspace{0.3\textwidth}
\subfigure[$~$]{\includegraphics[width=0.6\linewidth]{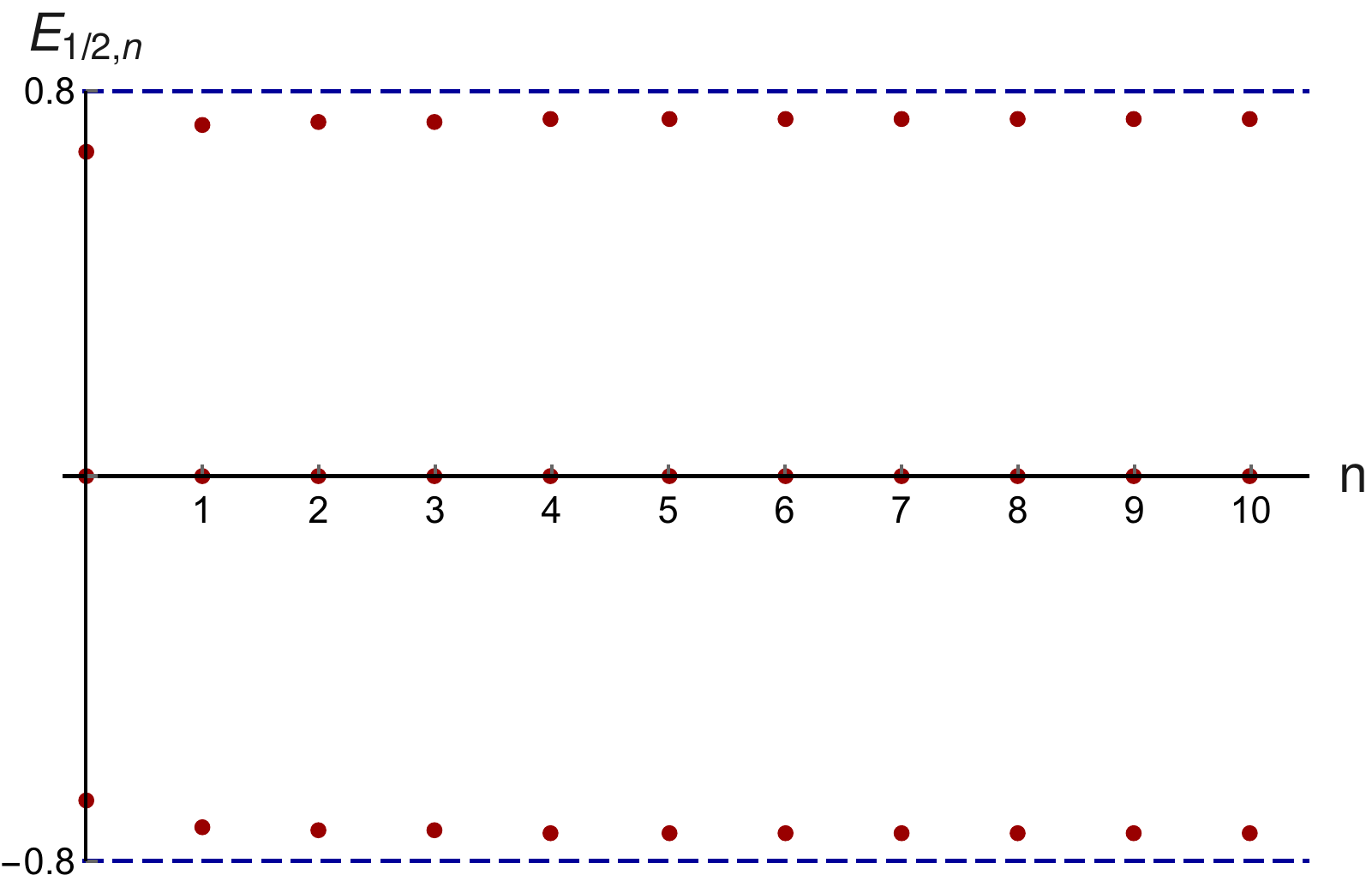}\label{fig6b}}\\ \hspace{2mm}%\hfill or \hspace{5mm} or \hspace{0.3\textwidth}
\subfigure[$~$]{\includegraphics[width=0.6\linewidth]{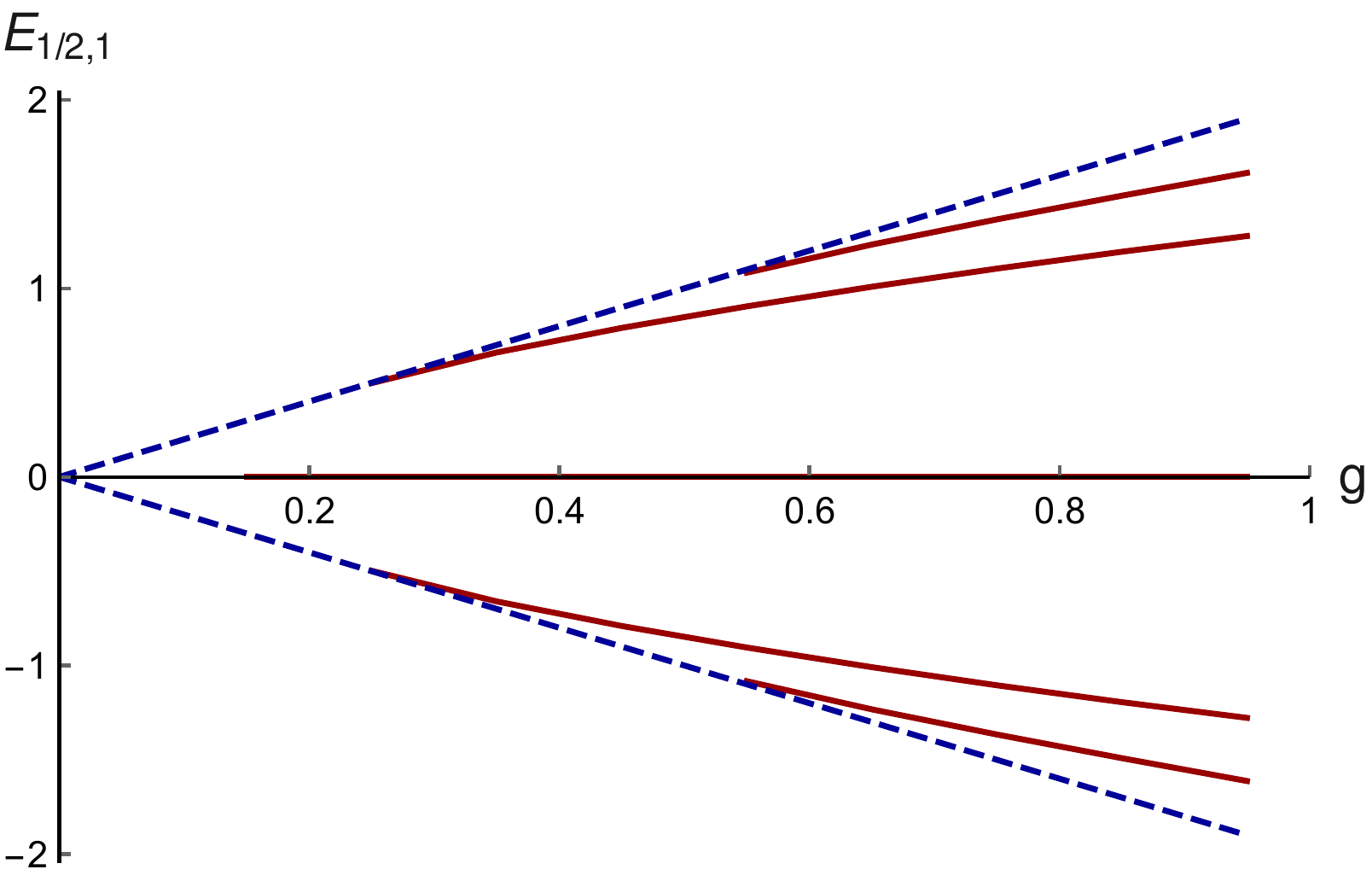}\label{fig6c}} %\hspace{5mm}
  	\caption{Fermion spectra for the third model, in terms of: (a) $\alpha$, for $g = 0.4$ and $n = 1$; (b) $n$, for $g = 0.4$ and $\alpha = 0.5$, and (c) $g$, for $\alpha = 0.5$ and $n = 1$.  \label{fig6}}
  \end{figure}

We now consider the third model, which is described by the function $f(\chi)=1/\sin^2\left((2n+1)\pi\chi\right)$ and $n=0,1,2,...$. Here the solution \cite{bazeia2020geometrically} is yet given by Eq. \eqref{kinkm2}, but now we have to use $y_n^{\pm} =(2 n+1) \left(1\pm \tanh(\alpha x)\right)$ instead. The kink profile is similar to the previous one, shown in Fig. \ref{fig3}; however, it now includes another plateau at its center, making the total number of plateaux to be an odd integer, counted as $2n+1$. In this case the fermion spectra also change, as displayed in Fig. \ref{fig6}.

%%%%%%%%%%%%%%%%%%%%%%%%%%%%%%%%%%%%%%%%%%%%%%%%%%%%%%%%%%%%%%%%%%%%%%%%%%%%%%%%%%%%%%%%%%%%%%%%%%%%%%%%%%%%%%%%%%%%
%%%%%%%%%%%%%%%%%%%%%%%%%%%%%%%%%%%%%%%%%%%%%%%%%%%%%%%%%%%%%%%%%%%%%%%%%%%%%%%%%%%%%%%%%%%%%%%%%%%%%%%%%%%%%%%%%%%%

%%%%%%%%%%%%%%%%%%%%%%%%%%%%%%%%%%%%%%%%%%%%%%%%%%%%%%%%%%%%%%%%%%%%%%%%%%%%%%%%%%%%%%%%%%%%%%%%%%%%%%%%%%%%%%%%%%%%
%%%%%%%%%%%%%%%%%%%%%%%%%%%%%%%%%%%%%%%%%%%%%%%%%%%%%%%%%%%%%%%%%%%%%%%%%%%%%%%%%%%%%%%%%%%%%%%%%%%%%%%%%%%%%%%%%%%%

%%%%%%%%%%%%%%%%%%%%%%%%%%%%%%%%%%%%%%%%%%%%%%%%%%%%%%%%%%
%%%%%%%%%%%%%%%%%%%%%%%%%%%%%%%%%%%%%%%%%%%%%%%%%%%%%%%%%%
\section{Discussion}

In this work we investigated the behavior of fermions in the background of kinklike structures that appeared recently in \cite{bazeia2020geometrically}. In the bosonic portion of the model, the field $\chi$ gives rise to a standard kink, which directly interferes in the profile of the field $\phi$, engendering an internal structure which depends on the choice of the function $f(\chi)$. This function encodes modifications on the elastic properties of the medium where the field $\phi$ evolves in the real line. The presence of the internal structure in the first model, which is shown in Fig. \ref{fig1}(a), is very much similar to the kinklike profile found before in Ref. \cite{rapid}, in the study of the magnetization in micrometer-sized ${\rm Fe}_{20}\rm{Ni}_{80}$ material in the presence of geometrical constrictions in the magnetic element. In this sense, we may say that the $\chi$ kink, together with the function $f(\chi)=1/\chi^2$, provides a way to modulate the geometrical constriction present in the magnetic element used in the experimental set up described in Ref. \cite{rapid}. As we have shown in the present investigation, the inclusion of fermions with the standard Yukawa coupling to the scalar field $\phi$, leads us to the fermion spectra displayed in Fig. \ref{fig2}(a,b). If we now recall that in the standard JR model, which can be obtained from the model here considered imposing that $\chi=\pm1$, the fermion spectrum only depends on $g$, and varies in a way similar to the case displayed in Fig. \ref{fig2b}. In fact, it contains the zero mode and some other massive states depending on the value of $g$. In this way, the first model is richer than the JR model, at least in the sense that the parameter $\alpha$ also contributes to the presence of massive bound states inside the gap in the fermionic spectrum, and this certainly modifies the fermion behavior in the presence of this new kinklike structure.

%%%%%%%%%%%%%%%%%%%%%%%%%%%%%%%%%%%%%%%%%%%%%%%%%%%%%%%%%%%%%%%%%%%%%%%%%%%%%%%%%%%%%%%%%%%%%%%%%%%%%%%%%%%%%%%%%%%%
\begin{figure}[t!]
 \subfigure[$~$]{\includegraphics[width=1.0\linewidth]{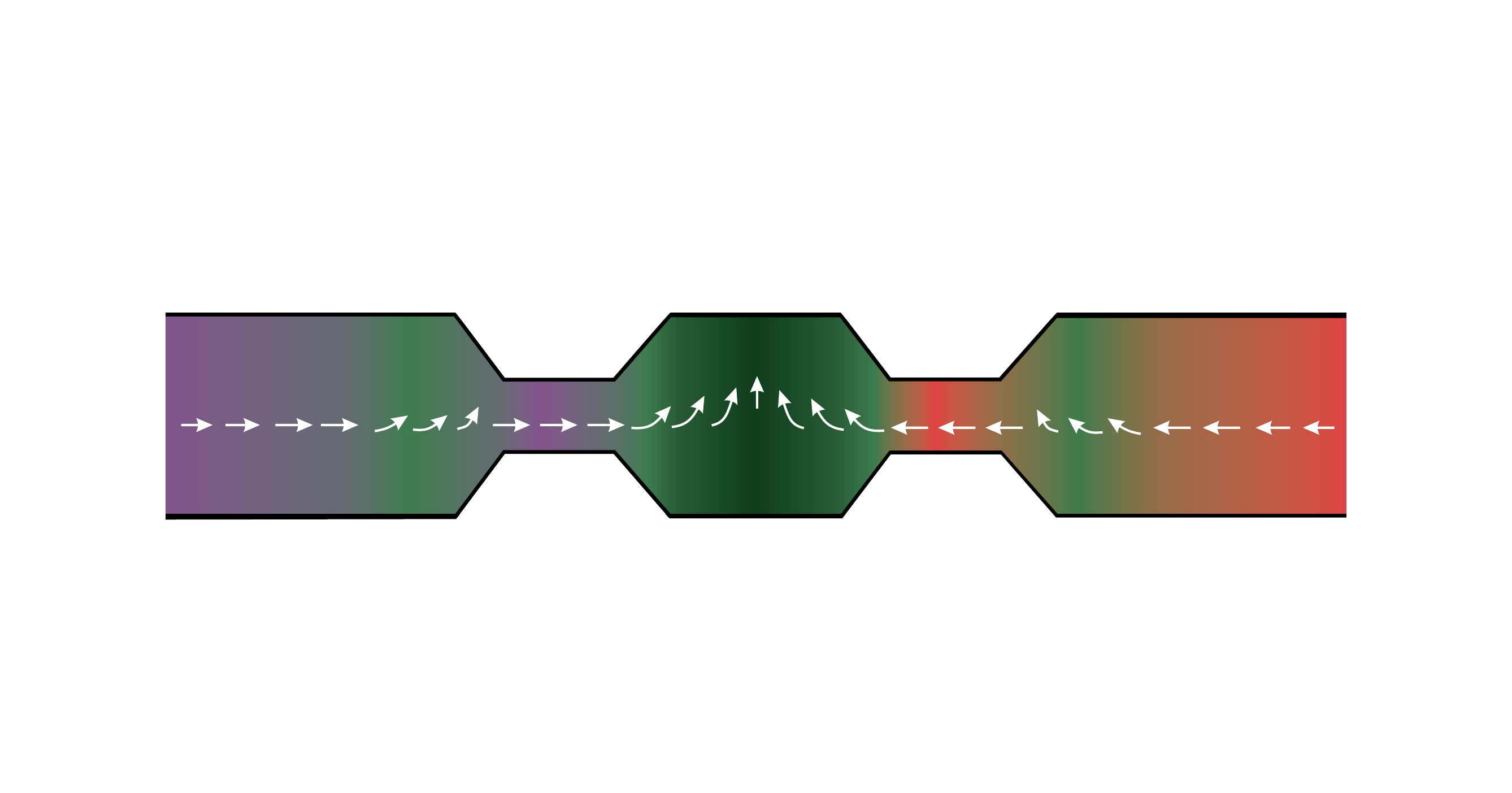}\label{fig7a}}\\ %\hspace{2mm}%\hfill or \hspace{5mm} or \hspace{0.3\textwidth}
\subfigure[$~$]{\includegraphics[width=1.0\linewidth]{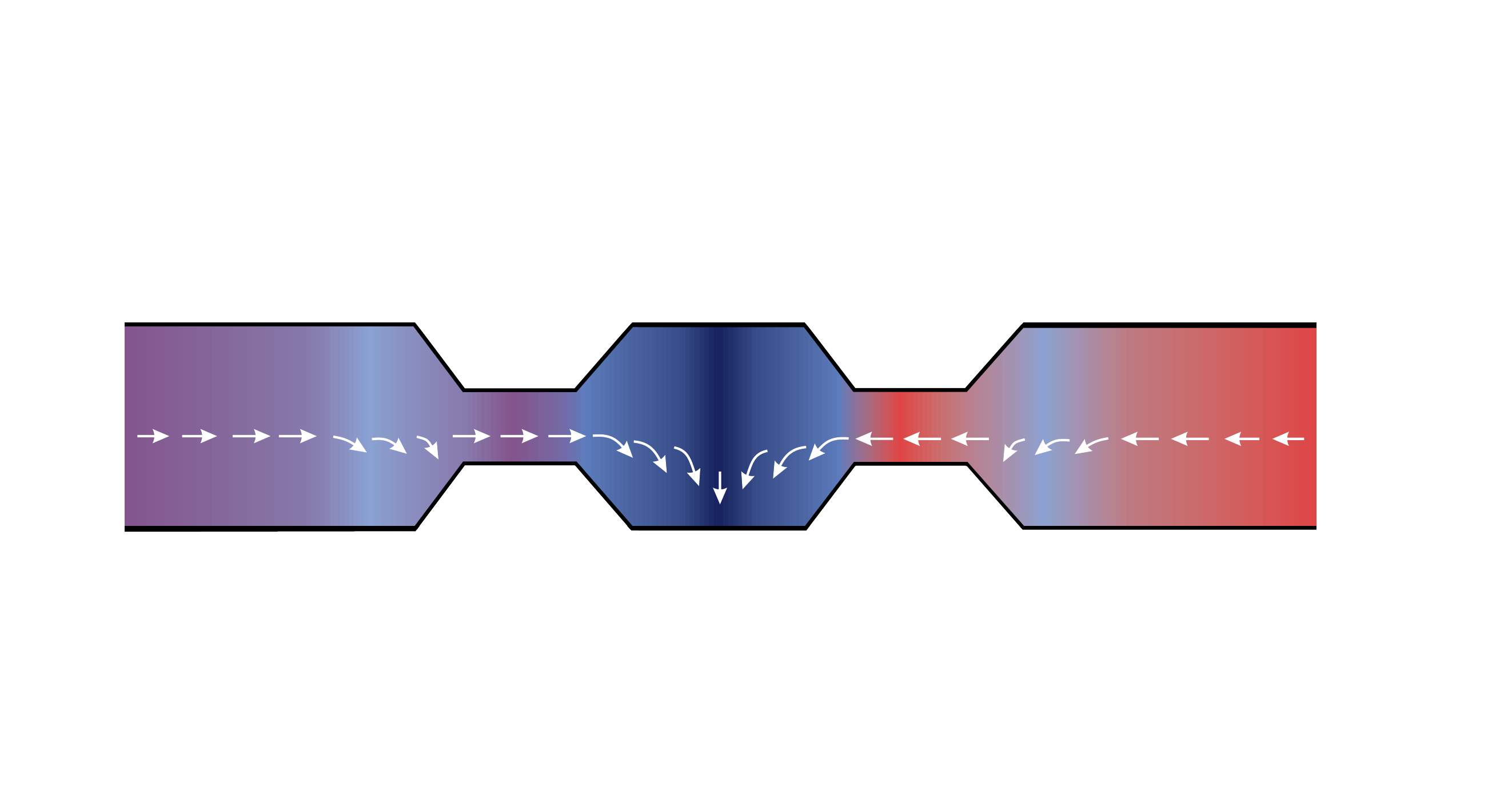}\label{fig7b}} %\hspace{5mm}%\hfill or \hspace{5mm} or \hspace{0.3\textwidth}
  	\caption{Illustration of the nanowire with the two geometric constrictions suggested from the second model with $n=1$, with the magnetization with the (a) up (dark green) and (b) down (dark blue) polarities. \label{fig7}}
  \end{figure}
%%%%%%%%%%%%%%%%%%%%%%%%%%%%%%%%%%%%%%%%%%%%%%%%%%%%%%%%%%%%%%%%%%%%%%%%%%%%%%%%%%%%%%%%%%%%%%%%%%%%%%%%%%%%%%%%%%%%
  
Inspired by the result of the first model, we decided to investigate two other possibilities, the second and third models, in which the internal structure of the kinklike configuration is richer than before. In the second model, for instance, the main effect comes from the presence of the new function $f(\chi)=1/\cos^2(n\pi \chi)$, and leads to a more involved kinklike profile which is depicted in Fig. \ref{fig3}(a,b). This modifies the spectra of fermions as shown in Fig. \ref{fig5}(a,b,c), in terms of the  parameters $\alpha, n, {\rm and}\,g$ which are present in this new model. Since the presence of massive states inside the fermion gap modifies the electronic behavior of electrons in the magnetic material, we then realised that an interesting study on the subject was already implemented experimentally in Ref. \cite{prlpolar}, on the control of domain wall polarity by current pulses. We notice that if we add geometric constrictions in a way similar to the case described in \cite{rapid}, in accordance with the results obtained above for the second model, we then expect that the presence of electric pulses will make the polarity of the wall to respond in a different way, and this would certainly modify the fermion response of such nanowires. In this sense, the above results seem to be of current interest to the area of nanoelectronics and may contribute to the fabrication of new electronic devices.

We now recall the experimental construction used in Ref. \cite{prlpolar} and think of a nanowire of the form shown in Fig. \ref{fig7}, with the light violet and light red standing for the magnetization pointing to the right and left along the horizontal axis, respectively. When the magnetization starts leaving the horizontal axis, it slowly starts becoming green or blue, if it changes to the positive or negative vertical axis, slowly going darker and darker. These two possibilities identify the two polarities of the wall structure. In this sense, the above results tell us that if we add the two geometrical constrictions also displayed in Fig. \ref{fig7}, the electronic current is modified by the presence of massive modes identified in Fig. \ref{fig5} and may contribute to change the mechanism for the inversion of polarity, unveiling another effect of current interest at the nanometric scale. This is in direct connection with Ref. \cite{rapid}: as we can read from its experimental results, the length and width of the constriction have approximately the same size of $500$ nm. Also, from Fig. 5 of \cite{rapid}, the plateaux in the kinklike configuration has the size which is also around $500$ nm. Thus, if one uses the second model, we see from the kinklike profile displayed in Fig. \ref{fig3} that it is possible to set $n=1$ and adjust $\alpha$ to determine the length, width and distance between the two constrictions which are illustrated in Fig. \ref{fig7}. The two parameters $\alpha$ and $n$ that appear in the second model seem to be adequate to put this model in good connection with the experimental implementation suggested in this work.

   The presence of geometric constrictions modifies the topological structure and changes the fermion behavior, adding distinct massive bound states to the spectrum of excitations, which may play a role in the construction of electronic devices at the nanometric scale. The theoretical result strongly suggests further experimental investigation, using techniques of direct imaging of the wall structure with high-resolution spin-polarized scanning electron microscopy \cite{rapid7} to understand the possibility of quantitatively measuring the importance and control of the new effect suggested in the present work. Due to the recent technological advancements \cite{rapid7,new,rmp}, the manipulation of magnetic domain walls in nanowires has increased considerably and may also foster new experimental studies in the subject.
   
%%%%%%%%%%%%%%%%%%%%%%%%%%%%%%%%%%%%%%%%%%%%%%%%%%%%%%%%%%%%%%%%%%%%%%%%%%%%%%%%%%%%%%%%%%%%%%%%%%%%%%%%%%%%%%%%%%%
\section*{Acknowledgments}

We would like to thank Victor Nunes for his help with the figures. This work has been partially financed by CNPq Grants Nos. 404913/2018-0 (DB), 303469/2019-6 (DB), and 305893/2017-3 (AM), Paraiba State Research Foundation Grant No. 0015/2019 (DB) and Universidade Federal de Pernambuco Edital Qualis A (AM).

\end{document}